\documentclass[prd,twocolumn,nofootinbib,preprintnumbers,superscriptaddress]{revtex4-2}
\usepackage{graphicx, epsfig}
\usepackage{color}
\usepackage{mathrsfs}
\usepackage{bm}
\usepackage{amsmath,amssymb}
\usepackage{mathrsfs}
\usepackage[caption=false]{subfig}
\usepackage[normalem]{ulem}
\usepackage{mathtools}
\usepackage[x11names]{xcolor}

\usepackage{physics}
\usepackage{amsthm}
\usepackage{soul}
\usepackage[colorlinks = true, linkcolor = purple, urlcolor  = blue, citecolor = blue, anchorcolor = blue]{hyperref}
\usepackage{float}
\usepackage[caption=false]{subfig}
\graphicspath{{images/}{./fig/}}

\definecolor{fashionfuchsia}{rgb}{0.96, 0.0, 0.63}
\colorlet{no_so_fashion_purple}{blue!50!red}

\newcommand{\be}{\begin{equation}}
\newcommand{\ee}{\end{equation}}
\newcommand{\ba}{\begin{eqnarray}}
\newcommand{\ea}{\end{eqnarray}}
\newcommand{\Eq}[1]{Eq.~(\ref{#1})}

\newcommand{\nn}{\nonumber}

\newcommand{\la}{\langle}
\newcommand{\ra}{\rangle}

\newcommand{\rh}{r_h}
\newcommand{\rhew}{r_{h,\rm EW}}

\newcommand{\rmcm}{{\rm cm}}
\newcommand{\rms}{{\rm s}}
\newcommand{\rmG}{{\rm G}}
\newcommand{\rmGeV}{{\rm GeV}}

\begin{document}

\preprint{NORDITA-2024-047}

\title{Spectra of magnetic fields from electroweak symmetry breaking}

\author{Tanmay Vachaspati}
\affiliation{Physics Department, Arizona State University, Tempe,  Arizona 85287, USA}

\author{Axel Brandenburg}
\affiliation{Nordita, KTH Royal Institute of Technology and Stockholm University, 10691 Stockholm, Sweden}
\affiliation{The Oskar Klein Centre, Department of Astronomy, Stockholm University, 10691 Stockholm, Sweden}
\affiliation{School of Natural Sciences and Medicine, Ilia State University, 0194 Tbilisi, Georgia}
\affiliation{McWilliams Center for Cosmology and Department of Physics, Carnegie Mellon University, Pittsburgh, Pennsylvania 15213, USA}

\begin{abstract}
We characterize magnetic fields produced during electroweak symmetry breaking
by non-dynamical numerical simulations based on the Kibble mechanism.
The generated magnetic fields were thought to have an energy spectrum $\propto k^3$ for small wavenumbers $k$,
but here we show that it is actually a spectrum $\propto k^4$ along with characteristic fluctuations in the magnetic helicity.
Using scaling results from MHD simulations for the evolution and assuming that the
initial magnetic field is coherent on the electroweak Hubble scale, we estimate 
the magnetic field strength to be $\sim 10^{-13}\, \rmG$ on kpc scales at the present 
epoch for non-helical fields. For maximally helical fields we obtain $\sim 10^{-10}\, \rmG$ 
on Mpc scales. We also give scalings of these estimates for partially helical fields.
\end{abstract}

\maketitle

\section{Introduction}
\label{Intro}

The standard model of particle physics implies that electroweak symmetry
breaking in the universe occurred at a temperature $\sim 100\, \rmGeV$ and at a 
cosmic time $\sim 10^{-10}\, \rms$ when the horizon size was $\sim 1\, \rmcm$.
It is not clear if the dynamics of electroweak symmetry breaking occurs as a first- 
or second-order phase transition or as a crossover, as this is sensitive to new
ingredients present in particle physics (dark matter, generation scheme for neutrino masses, etc.).
What is certain, however, is that the Higgs field responsible for electroweak
symmetry breaking acquires a non-vanishing vacuum expectation value (VEV).
It has been argued that the very process of the Higgs acquiring a VEV
generates a primordial magnetic field~\cite{Vachaspati:1991nm}, and
simulations of electroweak symmetry breaking by a few different groups~\cite{DiazGil:2007dy,DiazGil:2008tf,Zhang:2019vsb}
show that a few percent of the cosmic energy density at the electroweak
epoch is in the form of magnetic fields 
(see the reviews~\cite{Grasso:2000wj,Widrow:2002ud,Durrer:2013pga,Subramanian:2015lua,
Vachaspati:2020blt}).

Once electroweak symmetry has been broken in the universe, further evolution
of the magnetic field obeys Maxwell's equations in the presence of the cosmic
plasma. We expect the evolution to be well described by the equations of 
magnetohydrodynamics (MHD). However, the initial conditions for 
MHD evolution have to be derived from electroweak physics. Key features of the initial conditions are the energy and helicity spectra of the magnetic fields.
One approach to obtaining such initial conditions is to evolve the electroweak
equations through the process of electroweak symmetry breaking~\cite{DiazGil:2007dy,DiazGil:2008tf,Zhang:2019vsb}. These
field theory simulations are computationally expensive and their limited
dynamical range cannot resolve the spectral slope at small wavenumbers with sufficient degree of certainty.

In this paper we take a different approach to determine the spectra of magnetic
fields generated during electroweak symmetry breaking. The key idea is the
``Kibble mechanism'', also employed in studying the formation of topological
defects~\cite{Kibble:1976sj,Vilenkin:2000jqa}. The germ of the idea is that the Higgs VEV will 
be spatially (and temporally) varying during electroweak symmetry breaking.
Then, topological considerations, as discussed below, necessarily imply the presence of
magnetic charges and hence, magnetic fields. After electroweak symmetry
breaking has completed, the magnetic charges will have annihilated but the
magnetic field will survive. It is this magnetic field that we wish to characterize.
An advantage of using the Kibble mechanism is that it is not limited by dynamical 
range; a disadvantage is that it does not take into account any dynamics except 
for those dictated by symmetry considerations.

The calculation of magnetic fields resulting from the Kibble mechanism is
subtle because the algorithm necessarily produces magnetic monopoles
connected by Z-strings, also known as 
``Nambu dumbbells''~\cite{Nambu:1977ag,Patel:2021iik}.
A straightforward calculation of the magnetic field, ${\bf B}$, will not satisfy
${\rm div}\,{\bf B}=0$ and MHD evolution, as it assumes the absence of
magnetic charges, would not apply. Instead we want to construct the 
divergence-free magnetic field that results after the magnetic monopoles
have annihilated and the Nambu dumbbells have dissipated. We devise a
novel algorithm to construct the magnetic field, essentially by mimicking 
the eventual conversion of the Z-string to electromagnetic magnetic
field as seen in Ref.~\cite{Patel:2023ybi}.

We start in Sec.~\ref{ewsymmbreak} by defining the electromagnetic field 
strength in terms of electroweak variables. The topological aspects of the
standard model are outlined, together with the reason for the existence of 
magnetic monopoles. In Sec.~\ref{monopoles} we discuss the electroweak
magnetic monopole explicitly. This explicit monopole configuration is useful for testing the
numerical algorithm described in Sec.~\ref{algorithm}. An important feature
of our numerical algorithm is that it takes into account the Z-strings that,
when converted into electromagnetic fields, are necessary to ensure the 
divergence-free condition. The results of our numerical analysis
are described in Sec.~\ref{results}. In this section we also use current
results on MHD evolution to estimate the cosmological magnetic field at 
the present epoch. We conclude in Sec.~\ref{conclusions}. 
In Appendix~\ref{anotheralgorithm} we outline an alternate algorithm
to evaluate the magnetic field, one which accurately reproduces the
monopole configuration but is somewhat less convenient to implement in
the \textsc{Pencil Code} \cite{PC} that we use.
In Appendix~\ref{volumeavg} we revisit and correct the analytical estimate
in Ref.~\cite{Vachaspati:2020blt} to obtain the $k^4$ energy spectrum.
In Appendix~\ref{conversion} we relate some of our scaling exponents
to the symbols used earlier in the literature.

\section{Electroweak symmetry breaking and electromagnetism}
\label{ewsymmbreak}

In the standard electroweak model, the electromagnetic (as opposed to the Z-magnetic) magnetic field is given 
by~\cite{Vachaspati:1991nm}
\be
{\bf B} = \nabla \times {\bf {\cal A}} 
- i\frac{2\sin\theta_w}{g\eta^2} \nabla\Phi^\dag \times \nabla \Phi,
\label{Bfulldefn}
\ee
where ${\bf {\cal A}}$ is the electromagnetic gauge field, $\Phi$ is the vacuum
expectation value (VEV) of the Higgs field, $|\Phi |=\eta = 246\, {\rm GeV}$, and $\sin^2\theta_w =0.23$,
$g=0.65$ are coupling constants in the model. The last term in \eqref{Bfulldefn}
arises on requiring that the electromagnetic field strength should be gauge invariant
under electroweak symmetry transformations and should reduce to the usual
Maxwellian definition in ``unitary gauge'' in which $\Phi$ is constant.

The Higgs field is a complex doublet
\be
\Phi = \eta 
\begin{pmatrix}
\phi_1+i\phi_2 \\
\phi_3+i\phi_4
\end{pmatrix}
\ee
and the Higgs potential is
\be
V(\Phi) = \frac{\lambda}{4} ( |\Phi |^2 - \eta^2 )^2.
\ee
The vacuum manifold -- the minimum of the potential -- is given by
\be
|\Phi|^2 = \phi_1^2 + \phi_2^2 + \phi_3^2 + \phi_4^2 = \eta^2
\label{S3}
\ee
and this describes a three sphere, $S^3$. Symmetry dictates that the VEV
of the Higgs can take on any value on this $S^3$ with equal probability~\cite{Kibble:1976sj}. 
The VEVs in distant spatial regions will be independent of each other, implying
that $\nabla\Phi$ must be non-zero during electroweak symmetry breaking
and, from \eqref{Bfulldefn}, magnetic fields must be created~\cite{Vachaspati:1991nm}.

\begin{figure}
\includegraphics[width=0.50\textwidth,angle=0]{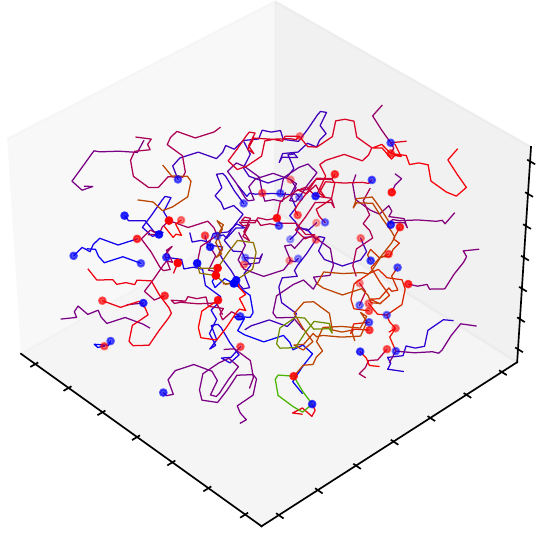}
 \caption{Distribution of monopoles and antimonopoles and the
 strings connecting them (from Ref.~\cite{Patel:2021iik}).
 }
 \label{mmbardistn}
\end{figure}

A more in-depth analysis reveals the topology in electroweak symmetry breaking~\cite{Gibbons:1992gt,Patel:2021iik}. 
Random orientations of $\Phi$ imply that the unit vector 
\be
{\hat n}^a = \frac{\Phi^\dag \sigma^a\Phi}{\eta^2}
\label{nvec}
\ee
will also be distributed randomly. The distribution of a three vector field can be
topologically non-trivial, implying the existence of Higgs zeros and magnetic
charges~\cite{tHooft:1974kcl,Polyakov:1974ek}. (The ``magnetic charge''
will be referred to as a ``magnetic monopole'' though it should be noted
that there is no static magnetic monopole solution of the electroweak equations.)
Further analysis shows that the electroweak magnetic monopole is connected
by a Z-string to an antimonopole and the Higgs field vanishes along the string
(see Fig.~\ref{mmbardistn})~\cite{Nambu:1977ag,Achucarro:1999it}. The dynamics 
of a monopole-antimonopole pair
shows that the system is unstable to decay and that the Z-string converts 
into electromagnetic magnetic field~\cite{Patel:2023ybi}. The final magnetic field,
after all the monopoles have annihilated, is divergenceless. We would like to determine 
the spectral properties of the final magnetic field on large length scales.

\section{Magnetic monopoles}
\label{monopoles}

The magnetic field in \eqref{Bfulldefn} has two pieces: the first term is
divergence free while the last term can have non-vanishing divergence
and represents the magnetic monopole contribution. We will be
interested in the spectral properties of the latter contribution to the
final magnetic field as it is insensitive to the detailed dynamics of electroweak
symmetry breaking. Then the magnetic field is given by
\be
{\bf B} = - i\frac{2\sin\theta_w}{g\eta^2} \nabla\Phi^\dag \times \nabla \Phi
\label{Bdefn}
\ee
and the corresponding gauge field using ${\bf B}=\nabla \times {\bf A}$ 
is\footnote{${\bf A}$ should not be confused with ${\bf {\cal A}}$ that occurs in
\eqref{Bfulldefn}.},
\be
{\bf A} = -i \frac{2\sin\theta_w}{g\eta^2} \Phi^\dag \nabla\Phi,
\label{Adefn}
\ee
where we have used $\nabla\times\nabla\Phi=0$.
As we shall see below, the gauge field defined in \eqref{Adefn} may
be multi-valued at certain points and it will be important to handle these
singular points carefully.

It is helpful to consider a specific configuration that corresponds to
an electroweak magnetic monopole. The Higgs field for the monopole can be written 
in spherical coordinates as
\be
\Phi_m = \eta \begin{pmatrix}
\cos(\theta/2) \\
\sin(\theta/2) e^{i\phi}
\end{pmatrix}
\label{monopolePhi}
\ee
for which the vector in \eqref{nvec} is,
\be
{\hat n}= (\sin\theta \cos\phi,\sin\theta\sin\phi, \cos\theta) = {\hat r}.
\ee
Then the magnetic field derived from \eqref{Bdefn} is
\be
{\bf B} =  \frac{\sin\theta_w}{g} \frac{\hat r}{r^2},
\label{monopoleB}
\ee
which is the magnetic field of a magnetic monopole.
The corresponding gauge field is
\be
{\bf A} = \frac{\sin\theta_w}{g} \frac{(1-\cos\theta )}{r\sin\theta} {\hat \phi},
\label{monopoleA}
\ee
where ${\hat \phi}$ is the unit vector in the azimuthal direction.
This is the gauge field of a magnetic monopole with a Dirac string at
$\theta=\pi$ where ${\bf A}$ is multi-valued. In the full electroweak model, 
$|\Phi|$ vanishes along the
Dirac string which gets replaced by a regular field configuration called
a Z-string. The Z-string carries Z-magnetic flux but no electromagnetic 
magnetic flux. However, the Z-string is unstable and rapidly decays by 
conversion into electromagnetic flux~\cite{Patel:2023ybi}. During this process the magnetic
monopoles annihilate and the magnetic field becomes divergenceless.

One way to study magnetic field production during electroweak symmetry
breaking is to numerically evolve
the dynamical electroweak equations, keeping track of the magnetic field
\cite{DiazGil:2008tf,DiazGil:2007dy,Zhang:2019vsb}.
However, such field theory simulations are computationally expensive 
and their dynamical range is limited. Recognizing that the 
monopole contribution
to the magnetic field arises simply due to variations of the Higgs field, it is
more efficient to perform a non-dynamical evaluation of the magnetic field,
though taking care to account for both the magnetic field of the magnetic 
monopoles and also the Z-strings that will eventually convert to electromagnetic 
magnetic fields.
We describe the algorithm for the non-dynamical evaluation in the next section.

\section{Algorithm}
\label{algorithm}

The magnitude of the Higgs field is fixed on the vacuum manifold: $|\Phi|=\eta$.
The direction of the Higgs field varies on a certain length scale that is determined
by the dynamics of electroweak symmetry breaking as in a first- or 
second-order phase transition, or a crossover. Then we imagine spatial domains
within which the direction of the Higgs is approximately constant, while the Higgs 
directions in different domains are completely uncorrelated. More concretely consider 
the Hopf parameterization of the Higgs field on its vacuum manifold,
\be
\Phi = \begin{pmatrix}
\cos\alpha \, e^{i\beta} \\
 \sin\alpha\, e^{i\gamma}
 \end{pmatrix},
 \label{hopf}
\ee
where $\alpha \in [0,\pi/2]$ and $\beta,\, \gamma \in [0,2\pi]$ are Hopf angles on
the three sphere given by \eqref{S3}. Electroweak
symmetry implies that the probability for $\alpha$, $\beta$ and $\gamma$ to
take on any value is given by the volume element $du\, d\beta\, d\gamma$ where
$u=\cos(2\alpha)/2$. Therefore $u \in [-1/2,1/2]$, $\beta \in [0,2\pi]$ and 
$\gamma \in [0,2\pi]$ are uniformly distributed in their respective domains.

We simulate the distribution of the Higgs field on a cubic lattice where each cell
of the lattice is considered to be a domain of constant Higgs -- the lattice
spacing corresponds to the domain size. Equivalently,
we assign a value of the Hopf angles to each vertex of a cubic lattice by
randomly selecting values from their probability distributions. The angles at 
any vertex define the Higgs field $\Phi$ at that vertex by \eqref{hopf}. 

In our numerical implementation, the zeros of the Higgs field
-- that are also the locations of the monopoles and Z-strings -- fall between 
lattice points and hence ${\bf A}$ is well-defined everywhere on the lattice. 
From ${\bf A}$
we find ${\bf B}$, not by taking the curl of ${\bf A}$, say by using finite
differences, but by calculating fluxes through plaquettes of the simulation
lattice,
\be
{\bf B}\, \delta x^2 = \hat p \oint_{\partial P} {\bf dl}\cdot {\bf A},
\label{BintA}
\ee
where $\partial P$ denotes the perimeter of the plaquette and we have 
assumed that ${\bf B}$ is spread uniformly over the area $\delta x^2$
of the plaquette and is oriented 
along the areal vector ${\hat p}$ of the plaquette (defined by the direction --
clockwise or counter-clockwise --
in which we perform the line integral). The advantage of using
\eqref{BintA} is that it automatically includes the contributions of the magnetic
monopoles and the strings. (In the electroweak model the strings are Z-strings 
that decay into electromagnetic magnetic fields; in electromagnetism the strings are
Dirac strings.)
The magnetic field that we obtain in this way will be divergenceless because
the magnetic monopole contribution to the magnetic field is compensated
exactly by the string contribution. We have tested our algorithm to make
sure that the string flux is $-2\pi$ so as to exactly cancel the $+2\pi$ flux of the
magnetic monopole.

A simple procedure to evaluate \Eq{BintA} is given by summing the
components of ${\bf A}$, defined by \eqref{Adefn},
along the links of each plaquette between
its four vertices I, II, III, and IV.
Starting at the lower left corner and going in the clockwise direction,
the contribution between vertices I and II yields
\ba
A_x|_a &=&-i\frac{2\sin\theta_w}{g\eta^2} \Phi^\dag\partial_x\Phi|_a
\nn \\
&\approx& -i \frac{\sin\theta_w}{g\eta^2 \delta x}
(\Phi^\dag_\textrm{II}+\Phi^\dag_\textrm{I})
(\Phi_\textrm{II}-\Phi_\textrm{I}),
\label{discretization}
\ea
where $\delta x$ is the lattice spacing.
Given that $\Phi^\dag_\textrm{II}\Phi_\textrm{II} =\eta^2$ and
$\Phi^\dag_\textrm{I}\Phi_\textrm{I} = \eta^2$ we are left with
\ba
A_x|_a &\approx&
-i \frac{\sin\theta_w}{g\eta^2 \delta x}(\Phi^\dag_\textrm{I}\Phi_\textrm{II}
-\Phi^\dag_\textrm{II}\Phi_\textrm{I})
\nn \\ 
&=& \frac{2 \sin\theta_w}{g\eta^2 \delta x} \textrm{Im}\,\Phi^\dag_\textrm{I}\Phi_\textrm{II}.
\ea
In this way, we obtain from all four contributions
\ba
&&
\oint_{\partial P} {\bf dl}\cdot {\bf A}\approx 
\nn \\ &&
\frac{2 \sin\theta_w}{g\eta^2}
\textrm{Im}\left(\Phi^\dag_\textrm{I}\Phi_\textrm{II}
+\Phi^\dag_\textrm{II}\Phi_\textrm{III}
+\Phi^\dag_\textrm{III}\Phi_\textrm{IV}
+\Phi^\dag_\textrm{IV}\Phi_\textrm{I}\right).
\label{BintA2}
\ea
We have checked that this algorithm gives the string flux of
$-2\pi$ so that it exactly cancels the $+2\pi$ flux of the magnetic monopole
at sufficiently high resolution.
It also agrees with an alternate algorithm presented in Appendix~\ref{anotheralgorithm}.
For convenience, we will work in units such that $2\sin\theta_w/ g = 1$, $\eta = 1$.

The above procedure evaluates the magnetic field flux through each plaquette
of the lattice, namely $B_x(i,j+1/2,k+1/2)$, $B_y (i+1/2,j,k+1/2)$
and $B_z (i+1/2,j+1/2,k)$ for all vertices $(i,j,k)$. Each component can be Fourier 
transformed using
\be
{\bf b}({\bf k}) = \int d^3x \, {\bf B}({\bf x} ) e^{+i{\bf k}\cdot {\bf x}}.
\ee
A translation of the coordinate system only introduces an overall
phase factor that does not enter the energy spectrum. For example, for the
$z$-component of the magnetic field we can shift the coordinate system by 
$(1/2,\,1/2,\,0)$ resulting in a magnetic field defined on a regular grid which
can be Fourier transformed using a Fast Fourier Transform routine. The
shift introduces an overall phase factor $\exp(i {\bf k}\cdot {\bf a})$ in the 
Fourier transform where ${\bf a}$ is the shift vector but this factor does not
affect the energy spectrum which only depends on $| {\bf b}({\bf k}) |^2$.
Further, the different components of the magnetic field enter the power
spectrum independently and can be evaluated using different suitable
shifts.

The magnetic field generated by the random distribution of $\Phi$
will be isotropic, homogeneous and divergenceless, and the two-point 
correlation function will take the standard form,
$\la b_i({\bf k}) b_j^*({\bf k}') \ra = (2\pi )^6 \delta^{(3)} ({\bf k}-{\bf k}') M_{ij}({\bf k})$
given by just two functions of the wavenumber $k$,
\ba
M_{ij}({\bf k})=\frac{E_M(k)}{4\pi k^2} p_{ij} + i\epsilon_{ijk} k^l \frac{H_M(k)}{8\pi k^2},
\label{corrfn}
\ea
where $p_{ij} = \delta_{ij}-{\hat k}_i {\hat k}_j$ is the projection operator
that ensures the divergenceless condition.
The function $E_M(k)$ is called the energy spectrum and $H_M(k)$ the
helicity spectrum. The energy spectrum is evaluated using,
\be
E_M(k) (2\pi)^3 \delta^{(3)}(0)  = \frac{k^2}{(2\pi)^2} \la | b_i({\bf k}) |^2  \ra .
\label{EMformula}
\ee
On the discrete lattice, the $(2\pi)^3 \delta^{(3)}(0)$ on the left-hand side is replaced 
by $V$, the lattice volume. Note that the form of the correlation function in~\eqref{corrfn}
only applies to divergenceless magnetic fields, that is after the magnetic monopoles and
connecting Z-strings have annihilated.

\begin{figure}
\includegraphics[width=0.50\textwidth,angle=0]{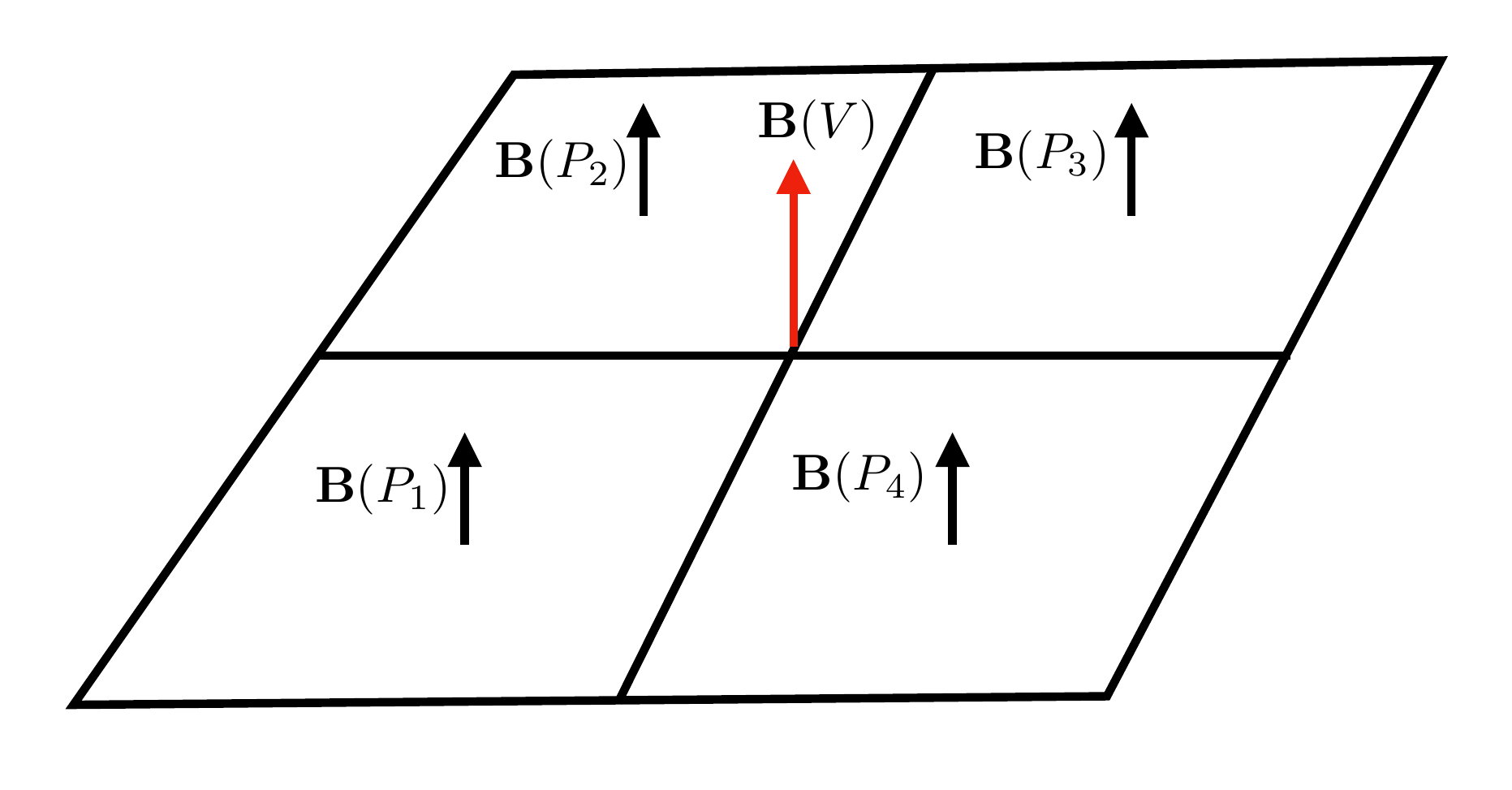}
 \caption{The $z$-component of the magnetic field at vertex $V$ is 
 determined by averaging the magnetic field components on
 the adjoining plaquettes: 
 $B_z (V) = (B_z(P_1)+B_z(P_2)+B_z(P_3)+B_z(P_4))/4$ where
 the four plaquettes $P_1,\ldots,P_4$ are in the $xy$-plane.
 }
 \label{interpolate}
\end{figure}

The scheme described above to find the
magnetic field is inadequate to evaluate the magnetic helicity 
${\bf A}\cdot {\bf B}$ as it only provides the gauge field component along the link 
at the link midpoint, and only the magnetic field component orthogonal to the 
plaquette and at the center of the plaquette. We would like to obtain all components
of the magnetic field at a point, say the vertices of the lattice. To do so, we
interpolate the magnetic field from four adjoining plaquettes, as illustrated
in Fig.~\ref{interpolate}. Once we have the magnetic field on the vertices of
the lattice, we Fourier transform to get ${\tilde {\bf b}}({\bf k})$ where the
tilde denotes that this is the interpolated magnetic field. Then
the gauge field Fourier coefficients are found using
\be
{\tilde {\bf a}}({\bf k}) = -i \frac{\bf k}{k^2} \times {\tilde {\bf b}}({\bf k}),
\ee
up to additive terms proportional to ${\bf k}$ that are omitted by our gauge choice.

The helicity spectrum is now evaluated using
\be
H_M(k) (2\pi)^3 \delta^{(3)}(0)  = 
\frac{k^2}{(2\pi)^2} \la {\tilde {\bf a}}({\bf k}) \cdot {\tilde {\bf b}}^*({\bf k}) \ra.
\ee
Since there is no parity violation present 
in the Higgs sector of the standard electroweak 
model\footnote{Parity violation in the electroweak fermionic sector may play a role 
in producing magnetic helicity but that is not accounted for in the present model.}, 
we will have $H_M(k)=0$
and only helicity fluctuations will be present. Hence we will also evaluate
the shell-integrated helicity variance spectrum,
\be
{\rm Sp}(h) = \frac{k^2}{8\pi^3 V} \oint d\Omega_k  |{\tilde h}|^2,
\ee
where $h={\bf A} \cdot {\bf B}$, ${\tilde h}$ is its Fourier transform,
and the integration is over solid angles in $k$-space.
The ``Hosking integral'', defined as,
\be
I_H = \frac{2\pi^2}{k^2} {\rm Sp}(h) \big |_{k\to 0}
\ee
is a conserved quantity in MHD evolution~\cite{HS21,HS22}.
Its gauge-invariance has been proven in Ref.~\cite{HS21}
and demonstrated numerically in Ref.~\cite{Zhou+22}.
As shown in those earlier papers, the decay of non-helical MHD turbulence is governed by
the conservation of $I_H$, while the decay of helical MHD turbulence
is governed by the conservation of the mean magnetic helicity density
$I_M\equiv\langle h\rangle$.

\section{Results}
\label{results}

\subsection{Spectral scaling at large scales}

\begin{figure}
\includegraphics[width=0.48\textwidth,angle=0]{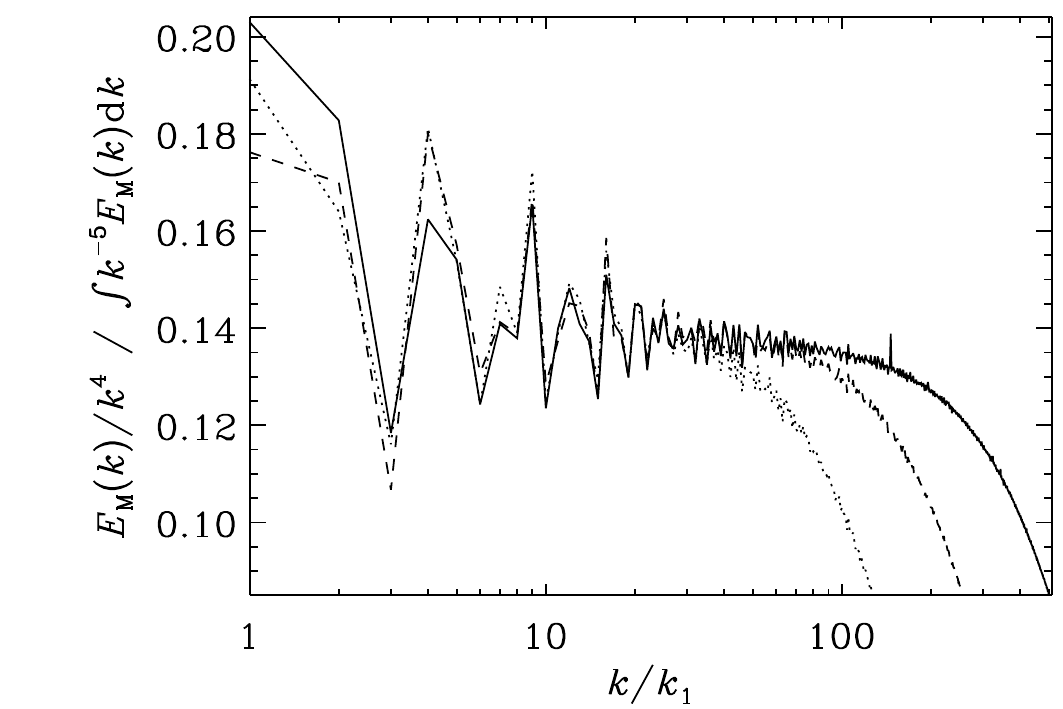}
 \caption{
Compensated magnetic energy spectra, $k^{-4}E_M(k)$, for $256^3$ (dotted line), $512^3$ (dashed line),
and $1024^3$ meshpoints (solid line), obtained by using \Eq{BintA2}.
All curves have been normalized by the value of $\int k^{-5}E_M(k)\,dk$ for the highest resolution case.
}
\label{EMfigure}
\end{figure}

We compute spectra from a three-dimensional mesh of size $L^3$, so the
smallest wave number in the domain is $k_1=2\pi/L$.
We use $N^3$ meshpoints, so the mesh spacing is $\delta x=L/N$ and
the largest wave number in the domain is the Nyquist wave number
$k_\mathrm{Ny}=2/\delta x=k_1\,N/2$.
Our results for the energy spectrum are derived using
\eqref{EMformula} and shown in 
Fig.~\ref{EMfigure}, compensated by $k^{-4}$ to show that, at large length scales, it gives the scaling
\be
E_M(k) \propto  k^4.
\label{scaling}
\ee
The exponent disagrees with the analytic argument in~\cite{Vachaspati:2020blt}
where the scaling was estimated to be $k^3$. In Appendix~\ref{volumeavg} we clarify
and correct that argument to also show consistency with the numerical scaling of $k^4$.
In Fig.~\ref{EMfigure}, the turnover at large $k$ is due to the discretization
error associated with the estimate of the derivative in \Eq{discretization}
for finite mesh spacing and shifts to higher $k$ when $N$ is increased.

To obtain an estimate for the energy spectrum and not just the scaling, we
need some input from dynamical 
simulations~\cite{DiazGil:2007dy,DiazGil:2008tf,Zhang:2019vsb}. The first input 
is that the energy density in magnetic fields after symmetry breaking is roughly
10\% of the total energy density. The second input is that the spectrum
in dynamical simulations is highly peaked at the largest coherence
scales in the simulations and suggests that the initial coherence scale
should be comparable to the cosmological horizon size at the electroweak
scale. Even if the coherence scale is initially sub-horizon, it will grow
on the eddy turnover time scale which is given by the coherence scale
divided by the Alfv\'en speed and will be short compared to the Hubble time.
This can lead to significant growth of the
initial coherence scale as discussed in~\cite{Vachaspati:2001nb}.
Using the connection between energy density and the energy spectrum,
\be
\rho_B = \int dk \, E_M(k),
\ee
we get an estimate for the magnetic energy spectrum immediately after 
electroweak symmetry breaking,
\be
E_M(k,t_{\rm EW}) \sim  \frac{\rho_{\rm EW}}{k_{\rm EW}} 
\left ( \frac{k}{k_{\rm EW}} \right )^4, \ \ k < k_{\rm EW},
\ee
where $\rho_{\rm EW} \sim (100\, {\rm GeV})^4 \sim (10^{24}\, {\rm G})^2$ is 
the energy density at the
electroweak scale, and $2\pi/ k_{\rm EW} \sim 1\, {\rm cm}$ is the physical
horizon size at the electroweak epoch.

\begin{figure}
\includegraphics[width=0.48\textwidth,angle=0]{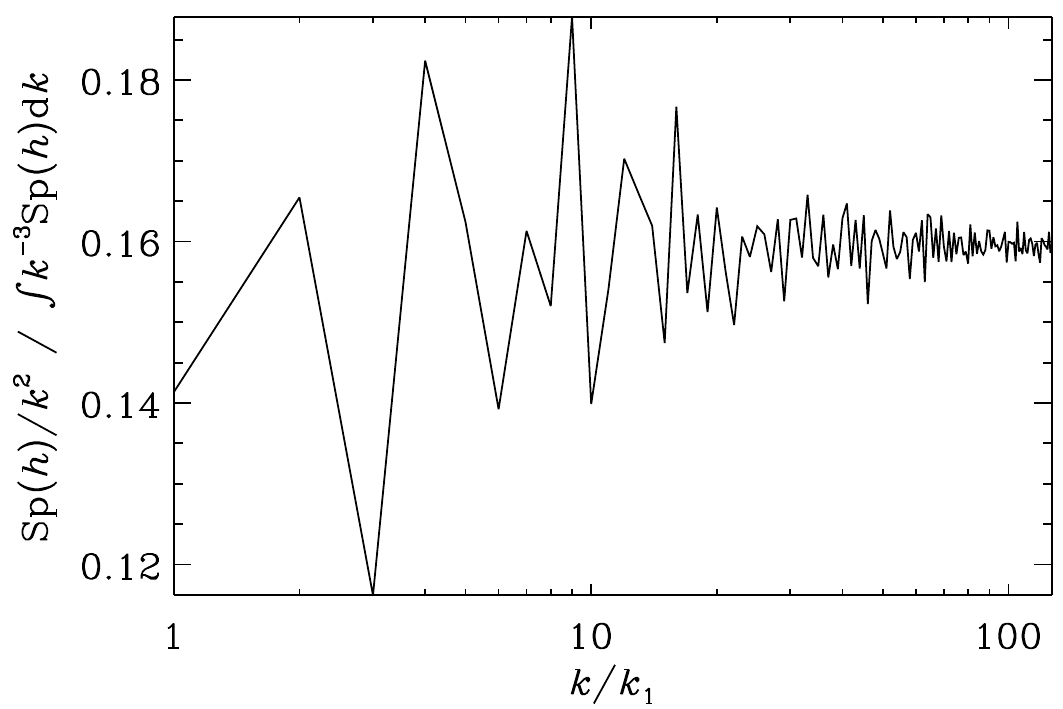}
 \caption{
Compensated shell-integrated helicity variance spectrum, $k^{-2}\mathrm{Sp}(h)$, using $N=256$ meshpoints,
normalized by the value of $\int k^{-3}\mathrm{Sp}(h)\,dk$.
}
\label{HM}
\end{figure}

We do not expect any net magnetic helicity in our simulations as there is no
source of parity violation and $H_M(k)=0$. In Fig.~\ref{HM} we plot the
$k^{-2}$ compensated shell-integrated helicity variance spectrum
and find it to be approximately constant on all scales.
Similarly to the normalization employed in the compensated spectrum
in Fig.~\ref{EMfigure}, we have normalized the compensated spectrum
$k^{-2}\mathrm{Sp}(h)$ by $\int k^{-3}\mathrm{Sp}(h)\,dk$.

\subsection{Inverse cascade scaling}

To estimate the magnetic field predicted from EWSB at the present epoch
we shall assume that the initial coherence scale is given by the cosmological
horizon size and has a $k^s$ spectrum for small $k$. We will give
estimates for general $s$.
Our numerical estimates for the location and amplitude of the
peak of the spectrum will turn out to be independent of the initial slope
of the subinertial ($k< k_{\rm peak}$) range, though other predictions
can be sensitive to the actual value of $s$.

\begin{figure}
\includegraphics[width=0.5\textwidth,angle=0]{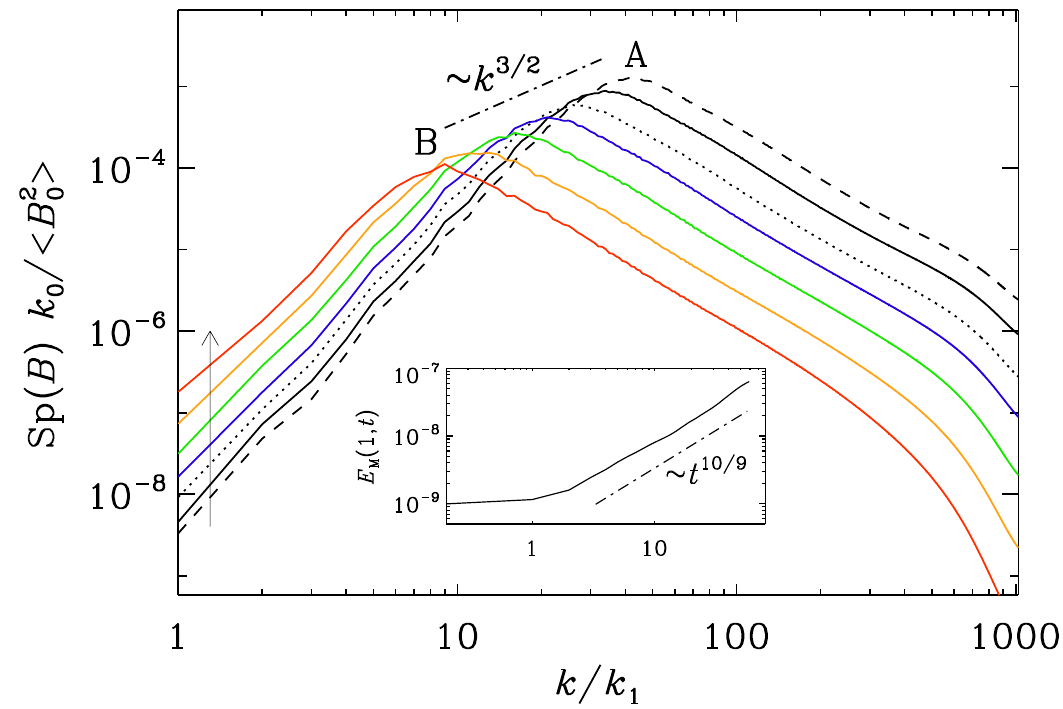}
 \caption{$E_M(k,t)$ vs $k$ for different times from MHD evolution of an
 initial $k^4$ spectrum with vanishing helicity.
 Point $A$ is the peak location at the initial time and point $B$ is the peak location 
 at the final time. The decay of the peak from $A$ to $B$ follows a $k^{3/2}$
 envelope. The vertical arrow and the inset show growth of the power 
 $\propto t^{10/9}$ on small $k$.
 }
 \label{mhdEvolution}
\end{figure}

As time evolves,  for the non-helical case the peak of the energy spectrum 
moves to lower $k$ as given by the $k^\epsilon$ ($\epsilon=3/2$) 
envelope in Fig.~\ref{mhdEvolution}, and the 
spectrum grows as $\tau^\gamma$, where $\tau$ is conformal time,
with~\cite{Brandenburg:2016odr, Brandenburg:2023rrd}
\be
\gamma = \frac{2(s-\epsilon)}{\epsilon +3}
\label{gamma}
\ee
for small $k$, as follows from
the scaling arguments of Ref.~\cite{Brandenburg:2016odr} and confirmed by the inset of 
Fig.~\ref{mhdEvolution} for $s=4$ when $\gamma=10/9$ (see
Appendix~\ref{conversion} for a conversion of the symbols used here to
those used in earlier literature; the symbol $\gamma$ used above is not
to be confused with the Hopf angle that is denoted by the same symbol).
We wish to relate the peak of the spectrum
at the present epoch $\tau_0$ (labeled as point B in Fig.~\ref{mhdEvolution}) to the peak of the
spectrum at the electroweak epoch $\tau_{\rm EW}$ (labeled as point A).
From the $k^\epsilon$ envelope we find
\be
E_M (k_B, \tau_0)  = E_M(k_A,\tau_{\rm EW}) \left ( \frac{k_B}{k_A} \right )^\epsilon
\label{ABenvelope}
\ee
and using the $k^s$ spectrum and the $\tau^\gamma$ growth we also find
\be
E_M(k_B,\tau_0)= E_M(k_A,\tau_{\rm EW}) \left (\frac{k_B}{k_A} \right )^s
\left ( \frac{\tau_0}{\tau_{\rm EW}} \right )^\gamma.
\label{ABspectrum}
\ee
Note that the evolution is in terms of comoving quantities, for example,
$k = a k^{\rm phys}$ and $B = a^2 B^{\rm phys}$ where $a$ is the
cosmic scale factor.
Also note that the scaling laws used above are only valid at late times, which we assume is given by $\tau \gg \tau_{\rm EW}$, when transients have died out (see Sec.~\ref{decaylaws}).

Dividing \eqref{ABenvelope} by \eqref{ABspectrum} and using
\eqref{gamma} we get
\be
k_B = k_A 
\left ( \frac{\tau_{\rm EW}}{\tau_0} \right )^{2/(\epsilon+3)}.
\label{kB}
\ee
Inserting this relation into \eqref{ABenvelope} gives,
\be
E_M(k_B,\tau_0) = E_M(k_A,\tau_{\rm EW}) 
\left ( \frac{\tau_{\rm EW}}{\tau_0} \right )^{2\epsilon/(\epsilon+3)}.
\label{EM}
\ee
The spectral slope $s$ does not appear in these relations,
except for the constraint that $s>\epsilon$.

The conformal magnetic field on scale $\lambda = 2\pi/k$ is given by 
${\cal B}_\lambda (\tau) = \sqrt{2kE_M (k,\tau)}$ and the physical magnetic field
is the conformal magnetic field multiplied by $T^2$ where $T$ is the cosmic
temperature. Therefore the physical peak magnetic field today is
\be
B_{B}^{\rm phys} (t_0) = B_A^{\rm phys} (\tau_{\rm EW}) 
\left ( \frac{\tau_{\rm EW}}{\tau_0} \right )^p
\left ( \frac{T_0}{T_{\rm EW}} \right )^2,
\label{Bt0}
\ee
where $p=(\epsilon+1)/(\epsilon+3)$
and the ``phys'' superscript denotes physical, not comoving, quantities.
The peak magnetic field strength at the electroweak epoch is
$B_A^{\rm phys}(\tau_{\rm EW}) \sim 10^{24}\, {\rm G}$.

The conformal time is given by $\tau \propto 1/T$ in the radiation dominated
universe (for $T \gtrsim T_{\rm eq} \sim 1\, {\rm eV}$) and $\tau \propto 1/\sqrt{T}$ 
in the matter dominated universe. 
(We ignore the changes in the number of degrees of freedom and other
cosmological events for these simple estimates.)
The electroweak temperature is $T_{\rm EW} \sim 100 \, {\rm GeV}$
and the temperature at the present epoch is $T_0 \sim 10^{-4}\, {\rm eV}$. These
numbers give,
\be
\frac{\tau_0}{\tau_{\rm EW}}  = \frac{T_{\rm EW}}{T_{\rm eq}} 
\sqrt{ \frac{T_{\rm eq}}{T_0} } \ \approx 10^{13}.
\ee

For non-helical fields, when $I_H=\mathrm{const}$, we have $\epsilon=3/2$ and
\eqref{kB} and \eqref{Bt0} with $p=5/9$ 
give~\cite{Brandenburg:2017neh}
\be
k_B^{\rm phys} \sim (1\, {\rm kpc})^{-1}, \ \ 
B_{\rm kpc}^{\rm phys} (t_0) \sim 10^{-13} \, {\rm G},
\label{Bkpcnonhelical}
\ee
These estimates are independent of the slope of the spectrum $s$ which will only 
enter the estimate of the field strength on length scales other than the peak length 
scale by factors of $(k/k_B)^s$.

Our formulae in \eqref{kB} and \eqref{Bt0} can also be applied to the case of
maximally helical magnetic fields, when $I_M=\mathrm{const}$. Then $\epsilon=0$ because $E_M$
at the peak stays constant for maximally helical fields.
These numerical values give $p=1/3$ (see below \eqref{Bt0}) and the estimates are,
\be
k_B^{\rm phys} \sim (1\, {\rm Mpc})^{-1}, \ \ 
B_{\rm Mpc}^{\rm phys} (t_0) \sim 10^{-10} \, {\rm G},
\ee
independent of the spectral slope $s$.

\subsection{Partially helical decay}

The magnetic fields generated during EWSB are expected to be
partially helical~\cite{Cornwall:1997ms,Vachaspati:2001nb,Vachaspati:2021vam}.
How does partial helicity affect the evolution of the magnetic field and our
estimates? With partial helicity, the evolution in the early stages is as if
there was no helicity. Then $E_M$ follows Fig.~\ref{mhdEvolution} and
$\epsilon=3/2$. However, $k H_M$ is conserved and the field evolves
towards maximal helicity which is defined by $|H_M |_{\rm max} = 2 E_M/k$.
Once the field is maximally helical, it evolves with $\epsilon=0$.

Let us denote the conformal time at which the field becomes maximally helical 
by $\tau_*$. Then applying \eqref{kB} and \eqref{EM} from $\tau_{\rm EW}$ to $\tau_*$
with $\epsilon=3/2$,
and then from $\tau_*$ to $\tau_0$ with $\epsilon=0$, gives
\ba
k_B &=& k_A \left ( \frac{\tau_{\rm EW}}{\tau_*} \right )^{4/9} 
\left ( \frac{\tau_*}{\tau_0} \right )^{2/3}, \label{kh1}\\
E_M(k_B,\tau_0) &=& E_M(k_A,\tau_{\rm EW}) 
\left ( \frac{\tau_{\rm EW}}{\tau_*} \right )^{2/3}.
\label{EMh}
\ea

To estimate $\tau_*$, we first define the relative helicity,
\be
\rh (k,\tau) = \frac{k | H_M |}{2E_M}.
\ee
While $\rh$ is a function of $k$ and $\tau$ in general, the most relevant value
of $\rh$ is at the peak of the spectrum and we will only consider the relative
helicity at the peak scale $k=k_{\rm peak}$. 
Then $\tau_*$ is defined by $\rh(\tau_*)=1$.
Using \eqref{EMh} and the conservation of magnetic helicity gives,
\be
\tau_* = \tau_{\rm EW} 
\rhew^{-3/2}
\label{tau*}
\ee
where $\rhew$ is the helicity fraction at the electroweak epoch. 

The magnetic field becomes maximally helical prior to the present epoch if $\tau_* < \tau_0$,
that is, if $\rhew > ( \tau_{\rm EW}/\tau_0)^{2/3} \sim 10^{-9}$.
If, on the other hand, $\rhew < 10^{-9}$
then there isn't enough time for the field to become maximally helical and the
evolution is just as in the case of non-helical fields, as in \eqref{Bkpcnonhelical}.
If $\rhew > 10^{-9}$, Eqs.~\eqref{kh1} and \eqref{EMh} with \eqref{tau*} give,
\ba
&& k_B = k_A \rhew^{-1/3}
\left ( \frac{\tau_{\rm EW}}{\tau_0} \right )^{2/3},
\label{kh}\\
&&
E_M(k_B,\tau_0) = E_M(k_A,\tau_{\rm EW}) \rhew, 
\label{partialhestimate}
\ea
which give the numerical estimates,
\be
k_B^{\rm phys}  \sim (1\, {\rm Mpc})^{-1} 
\rhew^{-1/3}, \ \ 
B_B^{\rm phys} (t_0)  \sim 10^{-10} 
\rhew^{1/3} \ {\rm G}.
\ee

\subsection{Relation to earlier decay laws}
\label{decaylaws}
 
The scalings \eqref{kB} and \eqref{Bt0} are equivalent to earlier ones
in terms of length scales $\xi_M(\tau)\equiv k_B^{-1}$ and mean magnetic
energy densities ${\cal E}_M(\tau)\equiv B_B^2(\tau)/2$; see, e.g.,
Refs.~\cite{Brandenburg:2023rrd,BNV24}.
Such relations were typically written in
the form
\be
\xi (\tau) = \xi(\tau_i) \left [ 1 + (\tau-\tau_i)/\tau_{\rm eddy} \right ]^{2/(\epsilon+3)},
\label{xiForm}
\ee
\be
{\cal E} (\tau) = {\cal E}(\tau_i) \left [ 1 + (\tau-\tau_i)/\tau_{\rm eddy} \right ]^{-2p},
\label{calEForm}
\ee
where $\tau_{\rm eddy}$ is some relevant eddy turnover time.
Different variations of these expressions have in common that the late-time behavior
($\tau\to\infty$) can be written as
\be
\xi (\tau) = \left[\xi(\tau_i) / \tau_{\rm eddy}^{2/(\epsilon+3)}\right]\, \tau^{2/(\epsilon+3)},
\ee
\be
{\cal E} (\tau) = \left[{\cal E}(\tau_i) \tau_{\rm eddy}^{2p}\right]\, \tau^{-2p}.
\ee
In the case when $I_H$ is conserved with $2/(\epsilon+3)=4/9$ and
$2p=10/9$, this can also be written as
\be
\xi (\tau) = C_\xi^H I_H^{1/9} \, \tau^{2/(\epsilon+3)},
\ee
\be
{\cal E} (\tau) = C_{\cal E}^H I_H^{2/9} \, \tau^{-2p}.
\ee
Conversely, when $I_M$ is conserved with $2/(\epsilon+3)=2/3=2p=2/3$,
these relations can also be written as
\be
\xi (\tau) = C_\xi^M I_M^{1/3} \, \tau^{2/(\epsilon+3)},
\ee
\be
{\cal E} (\tau) = C_{\cal E}^M I_M^{2/3} \, \tau^{-2p}.
\ee
This may look like a useless introduction of new parameters, but the
important point here is that $I_H$ and $I_M$ are 
conserved quantities 
and the coefficients $C_i^j$ for $i=\xi$ or ${\cal E}$ and $j=H$
or $M$ are believed to be universal ones.
The coefficients are known from earlier work and are independent of
the initial conditions, while $I_H$ and $I_M$ depend on the initial
conditions but do not change during the subsequent evolution.
Thus, in reality, there are no free adjustable parameters in this expression anymore.

\subsection{Relevance of the prefactors}

Although the scalings \eqref{kB} and \eqref{Bt0} strictly apply to late times, they can be extended to arbitrary times,
but we must ensure that the starting values, $k_A$ and $B_A$, are physically meaningful.
This requires that the time since the beginning of the decay is at all
times a certain fraction $C_M$ of the Alfv\'en time,
$\tau_{\rm A}=\xi/v_A$, where $v_A=(2{\cal E}/\rho)^{1/2}$ is the Alfv\'en speed \cite{BJ04}.
One may therefore expect $\tau_{\rm A}= C_M (\tau-\tau_i)$.
However, as discussed in Ref.~\cite{HS22}, the correct counting of time
is uncertain, particularly at early times, and becomes more certain only
at late times.
Furthermore, the factor $C_M$ depends on the Lundquist number, but is
expected to reach an asymptotic value of about 50 \cite{BNV24}.
The correct prefactors in the scaling relations \eqref{kB} and
\eqref{Bt0} depend on which conserved quantity governs the decay.
When the decay is governed by $I_H=\mathrm{const}$, we have
\be
k_i=C_k^H I_H^{-1/9}\tau_i^{-4/9},\quad
B_i=C_B^H I_H^{1/9}\tau_i^{-5/9},
\label{IHexpr}
\ee
while when the decay is governed by $I_M=\mathrm{const}$, we have
\be
k_i=C_k^M I_M^{-1/3}\tau_i^{-2/3},\quad
B_i=C_B^M I_M^{1/3}\tau_i^{-1/3}.
\label{IMexpr}
\ee
Here, the subscript $i$ refers to the points $A$ or $B$, or to any other
moment in time.
The values of the prefactors $C_k^H$, $C_B^H$, $C_k^M$, and $C_B^M$
are still somewhat uncertain, but it is suspected that they are universal.
Furthermore, they were given in terms of length scales and energy
densities; see Table~\ref{Tprefactors} for a comparison showing that
different measurements resulted in similar values.

\begin{table}[htb]\caption{
Comparison of the prefactors $C_k^H$, $C_k^M$, $C_B^H$, and $C_B^M$ found in earlier work.
}\vspace{12pt}\centerline{\begin{tabular}{lllll}
prefactor & \cite{BL23} & \cite{Brandenburg:2023rrd} & \cite{BB24} \\
\hline
$C_k^H=1/C_\xi^H$            & 1/0.15              & 1/0.12              & 1/0.14              & $\approx7.1$ \\
$C_k^M=1/C_\xi^M$            &   ---               &  ---                & 1/0.12              & $\approx8.3$ \\
$C_B^H=\sqrt{2C_{\cal E}^H}$ & $\sqrt{2\times3.8}$ & $\sqrt{2\times3.7}$ & $\sqrt{2\times4.0}$ & $\approx2.8$ \\
$C_B^M=\sqrt{2C_{\cal E}^M}$ &   ---               &  ---                & $\sqrt{2\times4.0}$ & $\approx2.8$ \\
\label{Tprefactors}\end{tabular}}\end{table}

The quantities $I_H$ and $I_M$ can be approximated in terms of their
dimensions, as has been done previously \cite{Zhou+22,BB24}.
Here, the magnetic field is always understood as being expressed as an Alfv\'en velocity.
It is then possible to cast equations \eqref{IHexpr} and \eqref{IMexpr}
in terms of $k_A(\tau_{\rm EW})$ and $B_A(\tau_{\rm EW})$.
However, given that the original expressions are based on potentially
universal prefactors and on conserved quantities that can in principle
be accurately determined, it is clear that the expressions \eqref{kB}
and \eqref{Bt0} cannot be applied to arbitrarily chosen starting values.

\section{Conclusions}
\label{conclusions}

We have shown that arguments based on symmetries of the electroweak vacuum 
manifold imply the production of magnetic fields that are largely independent of
the details of the symmetry breaking process. We have used these arguments to evaluate
the energy spectrum of the magnetic field and find $E_M \propto k^4$ on large length 
scales (small $k$). Without CP violation in the model, the average magnetic 
helicity vanishes but there are helicity fluctuations on all scales.
Together with earlier results from field theory simulations
of electroweak symmetry breaking and MHD simulations of cosmological magnetic field
evolution, we have estimated the magnetic field strength and coherence
scale at the present epoch. For non-helical fields, we find kpc coherence scales
and $10^{-13}\, \rmG$ field strength; for maximally helical fields we find Mpc
coherence scales and $10^{-10}\, \rmG$ field strength. For fractional helicity,
the estimates are between these extreme values. Such estimates are 
consistent with current upper and lower bounds on cosmological magnetic
fields~\cite{Durrer:2013pga,Vachaspati:2020blt}.

An important assumption in these estimates is that the initial coherence scale of
the magnetic field is 
given by the horizon size at the electroweak epoch. There is some supporting 
numerical evidence from field theory simulations that show that the spectrum
is peaked at the largest length scales in the simulations~\cite{DiazGil:2007dy,DiazGil:2008tf,Zhang:2019vsb}. It
would be reassuring to see the result hold up in bigger simulation volumes.
If the initial coherence scale is sub-horizon,
the sub-horizon dynamics should be taken into account and
the estimates should be rescaled
accordingly.

The limitations of our analysis should be pointed out.
The non-dynamical algorithm of Sec.~\ref{algorithm} is suitable for determining 
properties of the magnetic field that
are insensitive to the symmetry breaking dynamics. For example, the
method can give us the spectrum $k^4$ but it cannot give us the
coherence scale of the magnetic field which depends on the evolution during the symmetry
breaking process. The coherence scale may also evolve with time while the present 
algorithm is static and can at best provide a snapshot of the evolution.
Lastly, we have ignored the contribution of the ${\cal A}$ term in \eqref{Bfulldefn} 
whereas the full dynamics of electroweak symmetry breaking will include all contributions.

\acknowledgements
TV is grateful to Antonino Midiri, Teerthal Patel for discussions and to Chiara Caprini, 
Ruth Durrer and Alberto Roper Pol for motivating the present analysis.
We thank the Bernoulli Center, EPFL, Lausanne for hospitality during the 
``Generation, Evolution and Observations of Cosmological Magnetic Fields'' 
workshop.
TV thanks Tufts Institute of Cosmology for hospitality while this work was being done.
This work was supported by the U.S. Department of Energy, Office of High Energy 
Physics, under Award No.~DE-SC0019470.
A.B.\ was supported in part by the Swedish Research Council
(Vetenskapsr{\aa}det) under Grant No.\ 2019-04234, the National Science
Foundation under Grant Nos.\ NSF PHY-2309135 and AST-2307698, and the
NASA ATP Award 80NSSC22K0825.  We acknowledge the allocation of computing
resources provided by the Swedish National Allocations Committee at
the Center for Parallel Computers at the Royal Institute of Technology
in Stockholm.

\vspace{2mm}
Data availability---The source code used for the
simulations of this study, the {\sc Pencil Code},
is freely available from Ref.~\cite{PC}.
The simulation setups and data that support the findings
of this article are openly available~\cite{DATA}.

\appendix

\section{Another algorithm}
\label{anotheralgorithm}

Once we have the Higgs distribution, we can calculate the gauge field using
\eqref{Adefn}. The gauge fields are defined on the links of the lattice and
we will only need the component of the gauge field along the link. 
From \eqref{Adefn} and  \eqref{hopf} it follows
\be
{\bf A} = \frac{2\sin\theta_w}{g\eta^2} 
\left [ \cos^2\alpha \,  \nabla\beta + \sin^2\alpha \, \nabla\gamma \right ].
\label{Aabc}
\ee
Then ${\bf A}$ on a link is related to the Hopf angles at the endpoints of the 
link. As an example, consider the link from point $(i,j,k)$ on the lattice to the
point $(i+1,j,k)$. We first need to evaluate $\alpha$ at the central point
$(i+1/2,j,k)$. This is done using
\be
\Phi (i+1/2,j,k)=  \frac{\Phi(i,j,k) + \Phi(i+1,j,k)}{| \Phi(i,j,k) + \Phi(i+1,j,k) |}
\ee
and then 
\be
\cos\alpha = | \Phi_1 (i+1/2,j,k) |, \ \ 
\sin\alpha = | \Phi_2 (i+1/2,j,k) |,
\ee
where the subscripts 1 and 2 on $\Phi$ denote its upper and lower components.
Calculating the gradients of $\beta$ and $\gamma$ in \eqref{Aabc} requires some
care since the angles are defined on a circle. For example,
\be
\nabla_x \beta (i+1/2,j,k)= [[ \beta(i+1,j,k)-\beta(i,j,k) ]] / \delta x,
\ee
where $[[\cdot ]]$ means that the difference is 
taken to lie in the interval $[-\pi,\pi]$.
Operationally, if the difference $\beta(i+1,j,k)-\beta(i,j,k)$ is larger than $\pi$, then
we subtract $2\pi$, and if it is smaller than $-\pi$, we add $2\pi$. We have 
tested this algorithm for $\Phi_m$ given by \eqref{monopolePhi} and found excellent agreement with the analytical result in \eqref{monopoleA} 
even for relatively coarse lattices.
However, we observe no significant difference in the spectral properties of the magnetic field by using this method.

\section{Analytical reasoning}
\label{volumeavg}

In Ref.~\cite{Vachaspati:2020blt} an analytical argument was used to derive the
energy spectrum of the magnetic field resulting from electroweak symmetry breaking.
The conclusion was a $k^3$ spectrum. In this appendix we revisit the argument,
identify an error, and correct it to obtain a spectrum consistent with our 
numerical results.

We consider the volume-averaged magnetic field vector defined as
\be
B_{V,i} = \frac{1}{V} \int_V d^3x\, B_i({\bf x}),
\ee
where $V$ is the averaging volume.
On average, $B_{V,i}$ vanishes owing to isotropy.
However, the variance of $B_{V,i}$ does not vanish,
\be
B_V \equiv \la {\bf B}_{V}^2 \ra^{1/2}
= \left \la \left ( \frac{1}{V} \int_V d^3x\, {\bf B}({\bf x}) \right )^2 \right \ra^{1/2},
\ee
where $\la \cdot \ra$ denotes ensemble averaging.

\begin{figure}
\includegraphics[width=0.49\textwidth,angle=0]{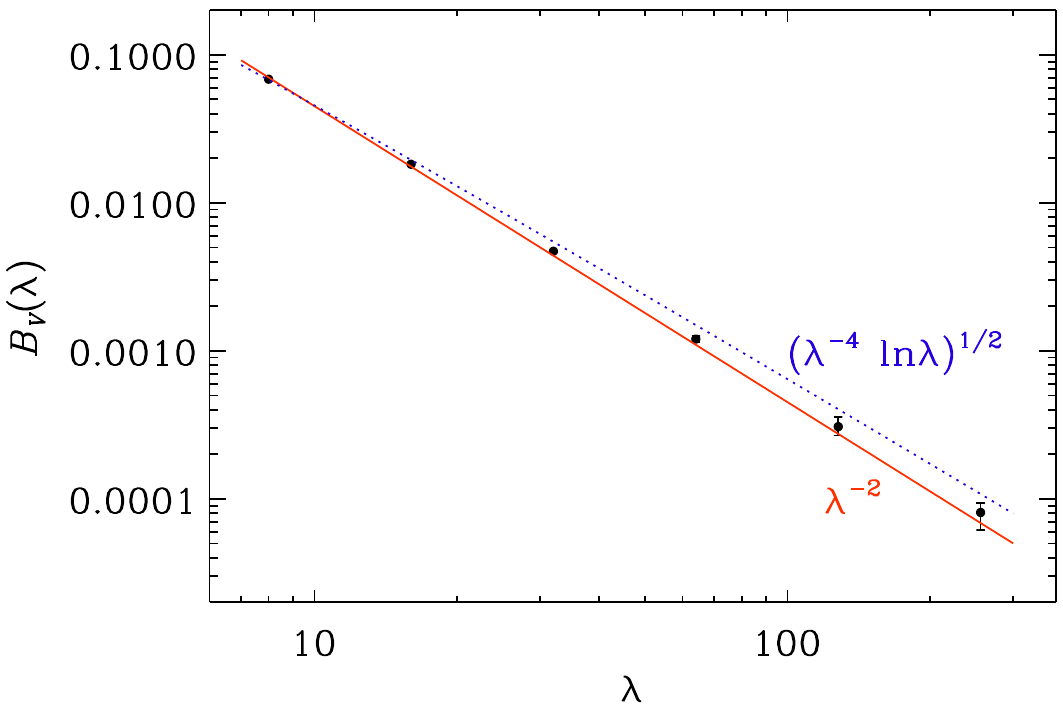}
 \caption{
Log-log plot of volume averaged magnetic field, $B_V$, versus size of volume shown as black dots.
The error bars have been computed based on averaging over the three coordinate directions of the magnetic field.
The red line has a slope of $-2$ and the blue dotted line shows $(\lambda^{-4}\ln\lambda)^{1/2}$.
}
\label{BV}
\end{figure}

To estimate $B_V$, we use Eq.~\eqref{Bdefn},
\ba
 \frac{1}{V} \int_V d^3x\, B_i ({\bf x}) &=&
 - i\frac{2\sin\theta_w}{g\eta^2 V}  \int_V d^3x\,  \epsilon_{ijk} \partial_j \Phi^\dag  \partial_k \Phi \nonumber \\
 &=&
 - i\frac{2\sin\theta_w}{g\eta^2 V}  \int_{\partial V} dS^j\,  \epsilon_{ijk} \Phi^\dag \partial_k \Phi, \nonumber \\ 
 \ea
 where $\partial V$ is the boundary of the volume $V$. The Higgs field magnitude is fixed to
 be the VEV, denoted by $\eta$, and so the gradients of $\Phi$ are of order $\eta /d$, where
 $d$ is the domain size over which $\Phi$ is approximately constant. Then the integral is
 a sum of terms of magnitude $\sim \eta^2/d$ but with random signs. If $V = \lambda^3$,
 the number of independent terms in the integral is $\lambda^2/d^2$ and, as for a random
 walk, the integral grows with $\lambda$ as $\sqrt{\lambda^2/d^2} \propto \lambda$. The
 prefactor scales as $1/V = 1/\lambda^3$, implying that $B_V$ scales as $1/\lambda^2$
 and hence
\be
B_V \propto \frac{1}{\lambda^2} \propto k^2.
\label{BVk2}
\ee
This scaling is confirmed by our numerical analysis and is shown in Fig.~\ref{BV}.

The next step is to connect $B_V$ to the energy spectrum $E_M(k)$. 
The connection is~\cite{Vachaspati:2020blt}
\be
B_V^2 = 2 \int dk\, E_M(k)\, W_V^2(k),
\label{BV2}
\ee
where the window function $W_V(k)$ is defined as
\be
W_V(k) = 3\,j_1(k\lambda)/(k\lambda),
\ee
where $j_1(x)=(\sin x-x\cos x)/x^2$ is the spherical Bessel function.

In Ref.~\cite{Vachaspati:2020blt} it was estimated
that $B_V^2 \sim k E_M(k)$ but this is
not correct as we now see.
Let us consider power law forms of $E_M(k)$ with a cut-off at $k_*$,
\be
E_M(k) = \begin{cases}
A \, k^n, & k \le k_* \\
0, & k >  k_*
\end{cases}.
\ee

Then \eqref{BV2} can be written as
\be
B_V^2 = \frac{2A}{\lambda^{n+1}} \int_0^{K_*} dq\, q^n 
\left [ \frac{3}{q^3} \left ( \sin q - q \, \cos q \right ) \right ]^2,
\ee
where $K_* \equiv k_*\lambda \gg 1$ because we are interested in the magnetic
field on large length scales. For $n \ge 3$, the integral is dominated by the upper limit
where $q \gg 1$. Then we can estimate,
\be
B_V^2 \sim \frac{9A}{\lambda^{n+1}} \int^{K_*} dq\, q^{n-4}
\propto \begin{cases}
\lambda^{-4} \,\ln\lambda, & n=3 \\
\lambda^{-4}, & n \ge 4
\end{cases}.
\ee
Since \eqref{BVk2} tells us that $B_V^2 \sim k^4 \sim \lambda^{-4}$, we must have $E_M(k) \propto k^n$ for
$n \ge 4$. In particular, $E_M(k) \propto k^3$ does not follow.

Note that $E_M \propto k^4$ is consistent with the claim in Ref.~\cite{Durrer:2003ja} even 
though our reasoning is non-dynamical and causality considerations are not relevant.

\section{Relation to earlier used exponents}
\label{conversion}

To facilitate the comparison between the scaling exponents used here
and those used in earlier papers, we list in Table~\ref{Tconversion}
the most important ones.
We also list their values that are relevant when the MHD decay
is governed either by $I_H=\mathrm{const}$ or by $I_M=\mathrm{const}$.

\begin{table}[htb]\caption{
Relation between the scaling exponents used here and those used in earlier literature,
along with their values when the MHD decay is governed either by $I_H=\mathrm{const}$ or by $I_M=\mathrm{const}$.
}\vspace{12pt}\centerline{\begin{tabular}{llcc}
present work     & earlier work & $I_H=\mathrm{const}$ & $I_M=\mathrm{const}$ \\
\hline
$s$              & $\alpha$ \cite{Brandenburg:2017neh, Olesen97} & \multicolumn{2}{c}{--- arbitrary ---}\\
$\epsilon$       & $\beta$ \cite{Brandenburg:2017neh} & 3/2 & 0 \\
$2/(\epsilon+3)$ & $q$ \cite{Brandenburg:2017neh} & 4/9 & 2/3 \\
$p$              & $q\kappa$ \cite{BNV24} & 5/9 & 1/3 \\
\label{Tconversion}\end{tabular}}\end{table}

The scaling exponent $2/(\epsilon+3)$ in \Eq{ABenvelope} characterizes
the growth of the characteristic length scale with time and was denoted in
Ref.~\cite{Brandenburg:2017neh} by $q$.
The scaling of the magnetic field with the characteristic length scale
was denoted in Ref.~\cite{BNV24} by $\kappa$ and is related to the
exponent $p=(\epsilon+1)/(\epsilon+3)$ introduced in \Eq{Bt0} such
that $p=\kappa q$.
The exponent $\gamma = 2(s-\epsilon)/(\epsilon+3)$ in \Eq{gamma}
agrees with the expression $(\alpha-\beta)q$ in Eq.~(2.9) of
Ref.~\cite{Brandenburg:2023rrd}.

The observation that the exponents $p$ and $2/(\epsilon+3)$ add up to
unity reflects the fact that the Alfv\'en time is proportional to the
actual (conformal) time $\tau$.
This is true regardless of whether the MHD decay is governed by
$I_H=\mathrm{const}$ or by $I_M=\mathrm{const}$.

\bibstyle{aps}
\bibliography{paper}
\end{document}